\documentclass[review]{elsarticle}

\usepackage{algcompatible}
\usepackage{algorithm}
\usepackage{multirow}
\usepackage{xcolor}
\usepackage{xspace}
\usepackage{comment}
\usepackage{caption}
\usepackage{subcaption}
\usepackage{amsmath}
\usepackage{lineno,hyperref}
\usepackage[T1]{fontenc}
\newcommand{\blue}[1]{\color{black}#1\color{black}}
\newcommand{\purple}[1]{\color{black}#1\color{black}}
\newcommand{\fig}{Fig.}

\journal{Journal of Future Generation Computer Systems}

\begin{document}

\begin{frontmatter}

\title{Heuristic-based Mining of Service Behavioral Models from Interaction Traces}

\author{Muhammad Ashad Kabir\corref{mycorrespondingauthor}}
\address{School of Computing and Mathematics, Charles Sturt University, NSW, Australia}
\cortext[mycorrespondingauthor]{Corresponding author}
\ead{akabir@csu.edu.au}

\author{Jun Han, Md. Arafat Hossain
        and Steve Versteeg}
\address{School of Software and Electrical Engineering, Swinburne University of Technology, Melbourne, Australia}

\begin{abstract}
 	\label{sec:abstract}
 	Software behavioral models have proven useful for emulating and testing software systems. Many techniques have been proposed to infer behavioral models of software systems from their interaction traces. The quality of the inferred model is critical to their successful use. While generalization is necessary to deduce concise behavioral models, existing techniques of inferring models, in general, overgeneralize what behavior is valid. Imprecise models include many spurious behaviors, and thus compromise the effectiveness of their use.
 	In this paper, we propose a novel technique that increases the accuracy of the behavioral model inferred from interaction traces. The essence of our approach is a heuristic-based generalization and truthful minimization. The set of heuristics include patterns to match input traces and generalize them towards concise model representations. Furthermore, we adopt a truthful minimization technique to merge these generalized traces. The key insight of our approach is to infer a concise behavioral model without compromising its accuracy. We present an empirical evaluation of how our approach improves upon the state-of-the-art specification inference techniques. The results show that our approach mines model with 100\% precision and recall with a limited computation overhead. 
 \end{abstract}




\end{frontmatter}


\section{Introduction}\label{sec:introduction}
Behavioral models of software system capture the behaviors that these systems exhibit at runtime. They play an integral role in software design, validation, verification, maintenance and testing~\cite{r1:autotest,r2:Pacheco:2005,r3:1007977,r4:4700316,r5:Ohmann,r6:Mariani:2011,r7:Hangal:2002,r8:Babenko:2009,r9:Gabel:2012,r10:Beschastnikh:2011,r11:6951474,r12:Ernst:1999}. For example, they can help to verify the software implementation~\cite{r6:Mariani:2011,r7:Hangal:2002,r8:Babenko:2009} and guide test case generation to improve test coverage~\cite{r1:autotest,r2:Pacheco:2005}. In software maintenance, behavioral models assist engineers to understand the complex objects and their interactions~\cite{r5:Ohmann}, as well as support debugging, fault detection, and detecting other anomalies~\cite{r3:1007977,r4:4700316}. In testing enterprise software systems, behavioral models can be used to synthesize  service response messages, and thereby emulating inaccessible services~\cite{arafat:saner, Du:2015:ITM:2693208.2693221}.

Despite its importance in the software development process, an accurate behavior model of the system under consideration is often missing or incomplete. Software systems often deviate from their early specifications as development and maintenance progresses. This deviation leads to inconsistencies between the implementation and the requirements. As a result, even if a software specification is available, it may not reflect the behaviors of the latest version of the software.

Numerous techniques have been proposed to automatically infer software behavioral models. These techniques either (1) infer a system's behavioral model (mainly as a finite state automata (FSA)) by generalizeing the observed invocation sequences found in its interaction traces (e.g.,~\cite{lo:2009,lo:2006a,lorenzoli:2008}) or (2) identify high level properties - declarative class and method invariants - by observing how the state (internal variable) of the system changes at runtime (e.g.,~\cite{beschastnikh:2011,beschastnikh:2013}). In this paper, similar to (1) above, we also focus on inferring a system's behavioral model (as a FSA) from its interaction traces by capturing valid invocation sequences. 

While existing research on FSA-based behavioral model inference has made substantial contributions, they still have significant limitations, i.e., their inferred models are often overgeneralized, accepting invalid interaction sequences. Many of these approaches \cite{lo:2006a,lo:2006b, lo:2009,lorenzoli:2008,reiss:2001, walkinshaw:2008,cook:1998} have adopted the kTail algorithm \cite{biermann:1972} with some extensions. Thus, all of them suffer from the overgeneralization problem related to kTail's state merging principle and heuristics. (1) In kTail, the state merging is based on local behavior of states, i.e., it merges two states only by looking \purple{at } \textit{k} future consecutive operations, leading to imprecision when states share their near future but differ in their distant future (and even when satisfying certain temporal invariants). (2) kTail starts with inter-trace state merging rather than intra-trace state merging, failing to spot generalization clues existed in a trace. Recently, a new technique has been proposed, called Synoptic \cite{beschastnikh:2011, schneider:2010}, which infers the behavioral model (in a form similar to FSA) from system interaction traces. Synoptic starts with a coarse initial model with gross overgeneralization, and then refines it by making it satisfying  three types of temporal invariants when such invariants are evident. As there are many other possible types of invariants, 
temporal invariants cannot guarantee the complete capture of system constraints. Thus, Synoptic also generate overgeneralized behavioral models, in particular when  the considered three temporal invariants are not all the necessary constraints that exist for the system under consideration.

As a consequence of over-generalization, the inferred system models include many spurious behaviors, \blue{which negatively impact on their usefulness for validating software behaviors. A study shows that the FSA-based model inference techniques are unable to infer precise models from the systems having complex behaviors~\cite{LO20122063}. As such, imprecise and overgeneralized models are inferred from the complex and large systems. Overgeneralized models are not useful in validating and verifying software behaviors as such models include numerous illegal behaviors.}

In this paper, we present a technique to infer more precise system behavioral models (in the form of FSAs)  by using a set of heuristics. It combines heuristic-based generalization and truthful minimization. We generalize each individual trace based on our insight that a trace can be thought of as a separate path in a finite state machine, and thus can provide an indication of a potential general behaviour of a system, which can be deduced with high precision based on additional evidences observed in other traces. The intuition for our heuristics is that a set of valid  operation sequences of a system observed in its traces exhibit evidence to generalize the system behavior. 
While the repetition in a FSA model matches infinite number of possible sequences of operations, we argue that for a real software system model, however, we can infer potential repetition behavior based on a number of recurrences of related operations with particular orders observed in system traces. As the count of recurrence value is higher, the stronger the evidence and thus enhance the precision. We provide a control on the level of evidence required by allowing the user to set the extent of recurrence. Finally, we combine the generalized trace-specific models by applying a truthful state merging (or minimization) technique that makes our model concise without introducing spurious behavior.

This paper makes three major contributions:
\begin{itemize}
    \item we propose a novel heuristic based approach to mine service behavioral model that resolves/reduces over-generalization problem
    \item a novel use of FSA minimization algorithm~\cite{hopcroft:2001} in the model generalization process
    \item we present an empirical evaluation of our approach comparing with existing approaches using a number of real-world services.
\end{itemize}

The rest of the paper is organized as follows. In Section \ref{sec:motiv-problem}, we analyse the over-generalization problem of mining service behavioral models in the context of state-of-the art approaches with a real-world example. Section~\ref{sec:specminerframework} presents SpecMiner framework -- our novel approach to resolve the problem. Experimental results on a number of services that compare the effectiveness of our approach with the existing approaches are reported in Section~\ref{sec:evaluation}. Section~\ref{sec:related work} reviews related work. Section~\ref{sect:threat} discusses the assumptions and limitations of our approach. Finally, Section~\ref{sec:conclusion} concludes this paper and highlights future work.    


\section{Motivating Example and Problem Analysis}
\label{sec:motiv-problem}

In this section, we first present a motivating example, then provide a description of the general research problem, and finally analyze the problem in the context of existing strategies that have been used in dealing with the problem.

\subsection{Motivating example}
\label{sec:background}

\begin{figure}[tb]
	\centering
	\includegraphics[width=1\textwidth]{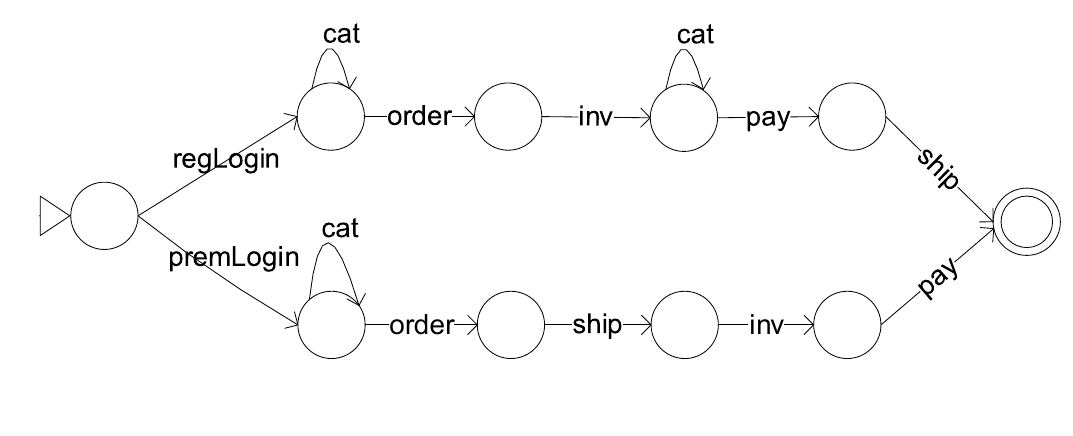}
	\caption{The specification of the \emph{retailer} service}
	\label{fig:retailerservice}
\end{figure}

Let us consider a Retailer service~\cite{4515865} that allows its clients to purchase products and supports two types of clients: regular and premium. A typical session or conversation of a regular client with the service may start with a request for the product catalog, followed by an order. Then, an invoice is sent to clients, and once the payment has been made, the requested goods are shipped. For a premium customer, goods may be shipped immediately after placing an order, and the invoice and payment can be done later \purple{on}). The behavior of the Retailer service is depicted in \blue{\fig}~\ref{fig:retailerservice}. In the rest of the paper, we use shortened forms of these operations or messages for simplicity: regLogin, cat, order, inv, pay, ship for regular customer and  premLogin, cat, order, ship, inv, pay for premium customer. 

\subsection{Problem description}
\label{sec:problem}
Given a set of system interaction traces, the problem of mining the behavioral model of the system from the traces can be stated as: \textit{infer from the traces a behavioral model which is as close as possible to the actual behavioral model of the system}~\cite{lo:2009}.

\begin{table}[b]
	\centering
	\caption{Example Traces}
	\label{tab:exampletraces}
	\begin{tabular}{cc}
		\hline 
		No.  & Trace  \\ \hline
		1 &  <regLogin,order,inv,pay,ship>  \\\hline
		2 & <premLogin,order,ship,inv,pay> \\\hline
		3 & <premLogin,cat,cat,order,ship,inv,pay> \\\hline
	\end{tabular}
\end{table}

\begin{figure}[tb]
	\centering
	\includegraphics[width=0.3\textwidth]{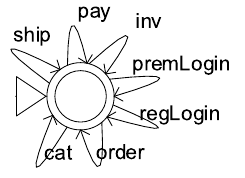}
	\caption{Flower model for \emph{retailer} service}
	\label{fig:flowermodelretailerservice}
\end{figure}

Let us consider the set of traces in Table~\ref{tab:exampletraces}. An extreme case of a concise and generalized model is illustrated in \blue{\fig}~\ref{fig:flowermodelretailerservice}. This model is overgeneralized because it accepts conversations that are absent in the actual system model and, in fact, accepts all possible conversations consisting of the retailer service messages. While only a subset of the possible conversations formed from these messages actually present in the actual system model (see \blue{\fig}~\ref{fig:retailerservice}).

\begin{figure}[tb]
	\centering
	\includegraphics[width=1\textwidth]{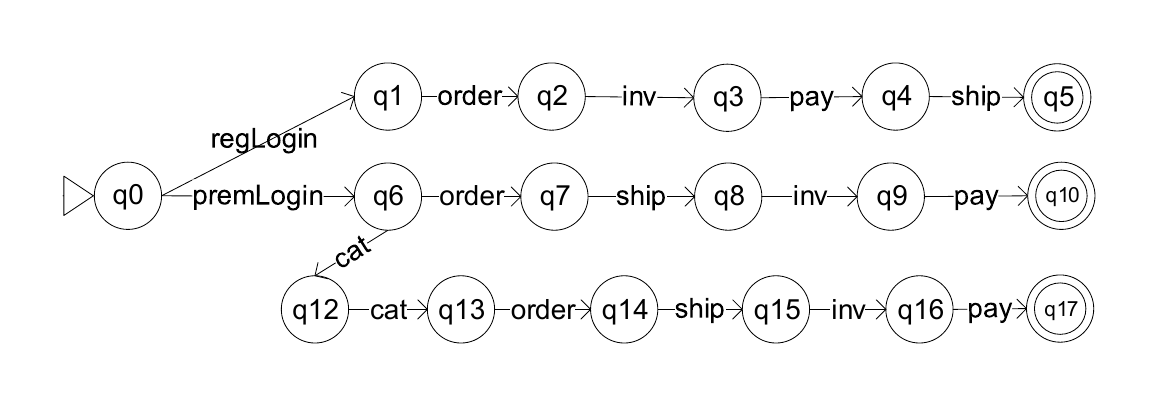}
	\caption{PTA model for example traces in Table \ref{tab:exampletraces}}
	\label{fig:ptamodelretailerservice}
\end{figure}

At the other extreme, it is possible to build a model with a path from each distinct conversation or trace in the trace set (also referred to as prefix tree acceptor (PTA)). A PTA is a tree representation of a set of traces. Each trace corresponds to a path in the tree. The PTA is obtained by merging the initial states of each trace into a single initial state, and from that state, merging all those states that are reached by the same invocation sequence. Traces that share the same prefix, share also a sub-path in the tree. \blue{\fig}~\ref{fig:ptamodelretailerservice} shows a PTA for the three traces in Table~\ref{tab:exampletraces}. Such a model would be difficult to understand, having too many states. Furthermore, it is too \emph{precise}, i.e., \emph{under-generalized} or \emph{no generalization}, since it accepts only the conversations in the trace set, which usually reflect only a partial behavior of the actual system or model. 

To date, two main strategies have been proposed in the literature to balance the trade-off between over-generalization and under-generalization (see section~\ref{sec:related work} for a detail literature review): \textit{k-equivalence state merging} and \textit{invariant (temporal) enhanced state merging}. We examine these two strategies and identify their limitations in the following two subsections.

\subsection{K-equivalent state merging}
\begin{figure}[tb]
	\centering
	\includegraphics[width=.8\textwidth]{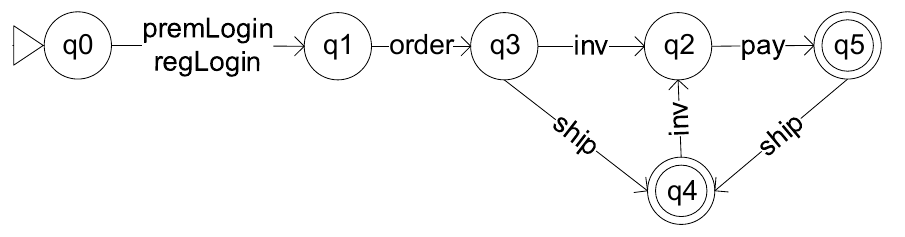}
	\caption{Inferred model using kTail (k=1) for trace1 and trace2}
	\label{fig:ktailwithk=1}
\end{figure}

\begin{figure}[tb]
	\centering
	\includegraphics[width=1\textwidth]{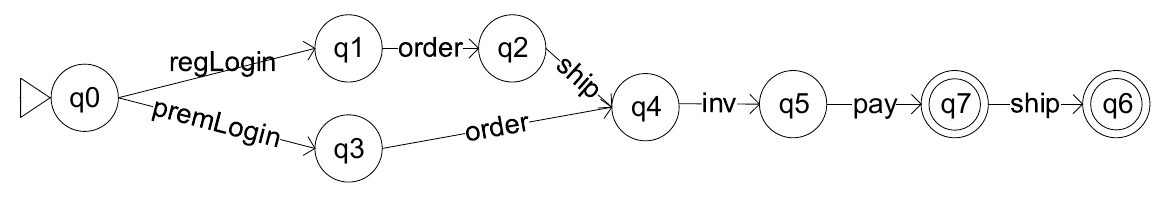}
	\caption{Inferred model using kTail (k=2) for trace1 and trace2}
	\label{fig:ktailwithk=2}
\end{figure}

The k-equivalent state merging, i.e., the kTail approach \cite{biermann:1972} captures the intuitive idea that if two states are not distinguishable by looking to their near future, they probably represent the same conceptual state, and thus can be merged. The kTail algorithm relies on selecting an appropriate \textit{k}, i.e., the number of  steps to be considered in distinguishing future states. This choice typically involves a trade-off between \textit{precision} (smaller \textit{k} implies more spurious merges due to the limited ``scope'', leading to \textit{overgeneralized}) and \textit{recall/completeness} (larger \textit{k} implies fewer merges, leading to \textit{undergeneralization}) of the generated model. For instance, if we apply the kTail algorithm with \textit{k}=1 on trace1 and trace2 of Table \ref{tab:exampletraces}, it merges state q3 with q9, q4 with q7, and q1 with q6, and infers an imprecise model shown in \blue{\fig}~\ref{fig:ktailwithk=1}. It allows many spurious behaviors, including \textit{<premLogin, order, inv, pay>} and \textit{<regLogin, order, ship>}, where a transaction completes without shipment and without payment, respectively. If we apply kTail with k=2, the algorithm merges state q2 of trace1 with q8 of trace2 of the PTA model in \blue{\fig}~\ref{fig:ptamodelretailerservice}. The resulting  model will be as shown in \blue{\fig}~\ref{fig:ktailwithk=2}. Even though the newly inferred 2-tail model is more precise than the 1-Tail model, it still allows spurious behaviors as shown in Table \ref{tab:spuriousmodel}.  

\begin{table*}[h]
	\centering
	\caption{Spurious behaviors of the inferred model using kTail (k=2)}
	\label{tab:spuriousmodel}
	\begin{tabular}{p{0.4\textwidth}|p{0.5\textwidth}}
		\hline 
		Trace  & Description  \\ \hline
		<premLogin, order, inv, pay> &  Transaction ended without shipment  \\\hline
		<regLogin, order, inv, pay> & Transaction ended without shipment \\\hline
		<premLogin, order, ship, inv, pay, ship> & Allows two shipments for a single order of premium customer \\\hline
	\end{tabular}
\end{table*}

\subsection{Invariant-enhanced state merging}
\begin{figure}[tb]
	\centering
	\includegraphics[width=.7\textwidth]{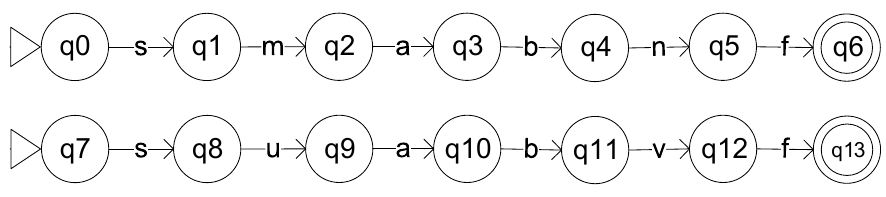}
	\caption{Example traces}
	\label{fig:kSteeringExampleTraces}
\end{figure}

\begin{figure}[tb]
	\centering
	\includegraphics[width=.75\textwidth]{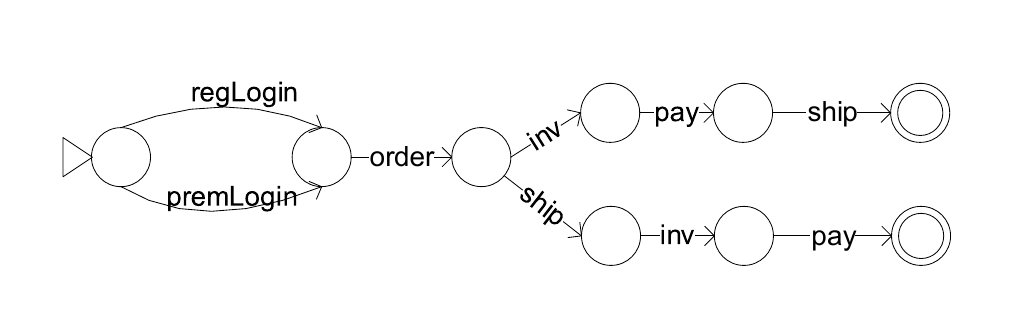}
	\caption{kSteering problem example}
	\label{fig:kstrprob}
\end{figure}

To reduce overgeneralization/imprecision, Lo et al. \cite{lo:2009} proposed kSteering that extends kTail by using temporal invariants inferred from traces to steer the generation of the FSA system behavioral model. The authors consider future-time and past-time temporal rules to prevent state merging that breaks non-local relations and thus improve the precision of the inferred FSA model. Future-time temporal rules, e.g., \purple{$e1\to e2$}, specify relations of the type ``whenever an event (say e1) happens (pre-condition) another event (say e2) must eventually happen (post-condition)''. Past-time temporal rules, e.g., \purple{$e1\gets e2$}, specify relations of the type `whenever an event (say e2) happens (pre-condition), another event (say e1) must have happened before (post-condition)".

kSteering prevents the merging of states across two traces if their future operation sets (i.e., the operation sets of these traces from after the state merging point to the end of the traces) are different from each other in terms of \emph{operation names}. Let us consider that \textit{<s, m, a, b, n, f>} and \textit{<s, u, a, b, v, f>} are two operation traces, as shown in \blue{\fig}~\ref{fig:kSteeringExampleTraces}. The kSteering can successfully prevents the merge of state q2 to q9 and q3 to q10, as they do not comply to the inferred temporal rules, \purple{$m\to n$, $m\gets n$, $u\to v$ } and \purple{$u\gets v$}. However, it can not handle spurious merges if the operation sets are same in terms of names but differ in sequencing. 
For example,  kSteering can not prevent the spurious merges between trace1 <regLogin, order, inv, pay, ship> and trace2 <premLogin, order, ship, inv, pay> (Table \ref{tab:exampletraces}) as the k-equivalence merges (k = 1, 2) do not violate the inferred temporal rules (see Table \ref{tab:inferredtemporalrulesksteering}). As a consequence, the inferred model (see Fig.~\ref{fig:kstrprob}), in this example, fails to capture the difference in service behavior between regular and premium customers. In particular, the model allows a purchase order from regular customer to be shipped before payment (this service should be allowed only for premier customer).

\begin{table}[h]
	\centering
	\caption{Inferred temporal rules by kSteering}
	\label{tab:inferredtemporalrulesksteering}
	\begin{tabular}{p{0.5\textwidth}|p{0.4\textwidth}}
		\hline 
		Future-time  temporal rules & Past-time temporal rules  \\ 
		 	\hline
		\purple{$regLogin\to order$, $regLogin\to inv$, $regLogin\to pay$, $premLogin\to order$, $premLogin\to ship$, $premLogin\to inv$, $premLogin\to pay$, $order\to inv$, $order\to pay$, $order\to ship$, $inv\to pay$} &  \purple{$order\gets inv$, $order\gets pay$, $order\gets ship$, $inv\gets pay$}
		\\\hline
	\end{tabular}
\end{table}

\section{The SpecMiner Framework}
\label{sec:specminerframework}

The goal of our approach is to infer a \emph{general}  behavioral model for a software system from its interaction traces with higher accuracy. An interaction trace is a sequence of operations and can be represented as a FSA with exactly one path. Thus, to generalize a set of traces into a concise system model, it is required to merge the states of their corresponding FSAs. To generate more accurate system model, we propose to use intra-trace state merging in contast to existing approaches such as kTail that use inter-trace state merging. Intuitively, a trace gives an insight into a system behavior. Thus, we argue that merging the states at a trace level will enhance the accuracy of generalization. 

\begin{figure}[htb]
	\centering
	\includegraphics[width=1\textwidth]{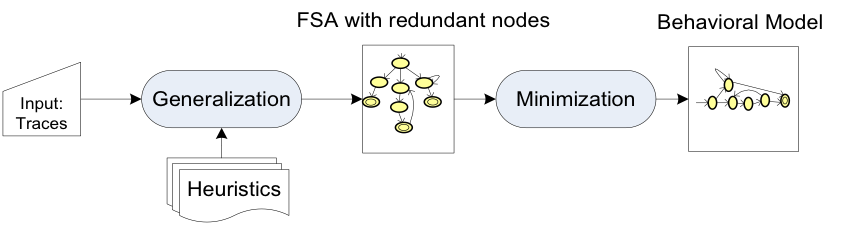}
	\caption{SpecMiner approach Overview}
	\label{fig:SpecMinerapproachoverview}
\end{figure}

As shown in \blue{\fig}~\ref{fig:SpecMinerapproachoverview}, our SpecMiner approach works in two phases: (1) heuristic-driven generalization -- merges states in traces to infer an automata that represents a generalize behavior of a service, and (2) automata minimization -- transforming an inferred automata that has a minimum number of states to make our generalize model deterministic and concise. 
We discuss the details of the two phases in the following subsections.

\subsection{Heuristic-driven generalization}
\label{subsec:heuric-driven generalization}
We propose a number of heuristics that serve as leads to search for evidences in traces for \blue{generalization}~and guide the generalization process. Heuristics are identified for each of the generalization properties (i.e., those related to iterations: cycle -- a path with a transition/edge that begins and ends at the same \purple{state}/vertex, loop -- a path with two or more connected edges that begins and ends at the same vertex, and multi-loop -- a special loop where a state has two or more loops) of the FSA model by analyzing their corresponding distinctive patterns that appear in traces. We discuss below the heuristics for each of the generalization properties, in turn.

\subsubsection{Loop}
\begin{figure}[htb]
	\centering
	\includegraphics[width=.7\textwidth]{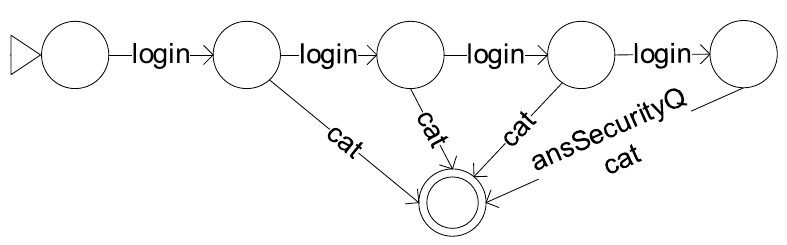}
	\caption{a partial system model that asks for security information after four incorrect login attempts}
	\label{fig:loginex}
\end{figure}

If a system model has a loop, we expect a trace with consecutive repetition of the loop operation. The repeat value/ frequency/count (RC) \textit{n} is an integer, where $n\geq2$. Repeat value 1 cannot be used to identify loop as it is similar to sequence. For most of the systems, repeat value 2 is enough evidence to precisely identify a loop, i.e., to overcome \blue{over-generalization}. However, there could be special cases, e.g., the login example presented in \blue{\fig }~\ref{fig:loginex} (where a system allows a maximum of four incorrect login attempts before blocking an account and requiring the user to answer security questions to proceed further). In this case, we need to set repeat count to 5. In general, it is recommended to set the repeat count as high as realistic and practical, but there should be at least one trace  with such a number of repeat loop operation occurrences.

\begin{figure}[htb]
\centering
\begin{subfigure}{1\columnwidth}
 \centering
    \includegraphics[width=.45\textwidth]{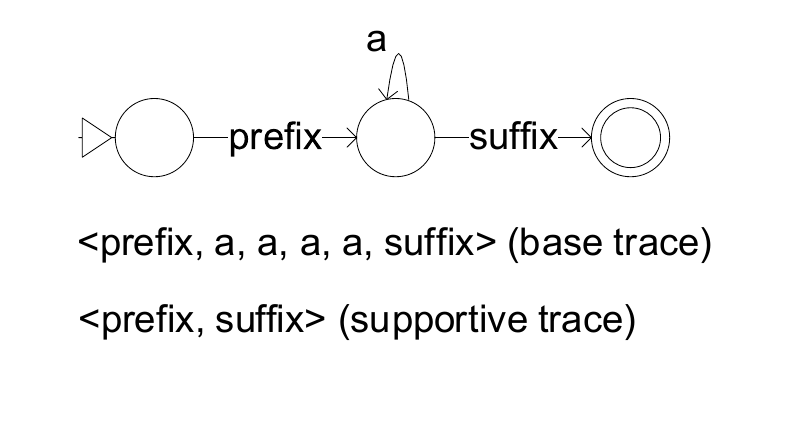}
    \caption{Case 1 (for repeat count (RC) = 2 or 3 or 4)}
    \label{fig:fig10a}
\end{subfigure}
\hfill
\begin{subfigure}{1\columnwidth}
 \centering
    \includegraphics[width=.5\textwidth]{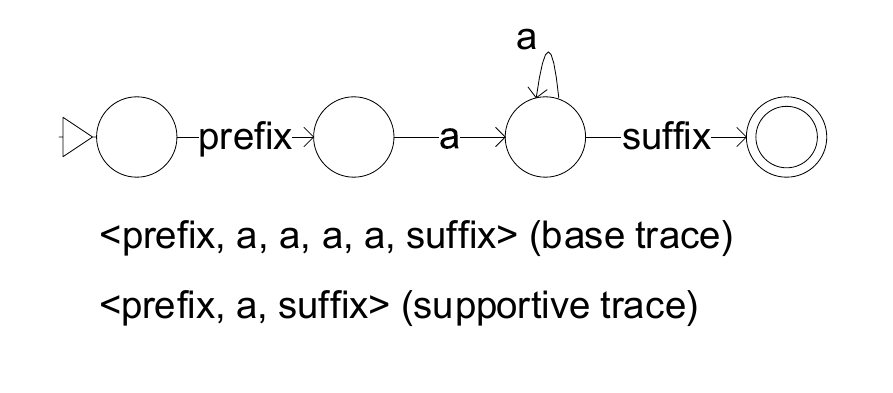}
    \caption{Case 2 (for repeat count (RC) = 2 or 3)}
    \label{fig:fig10b}
\end{subfigure}

\begin{subfigure}{.75\columnwidth}
 \centering
    \includegraphics[width=.8\textwidth]{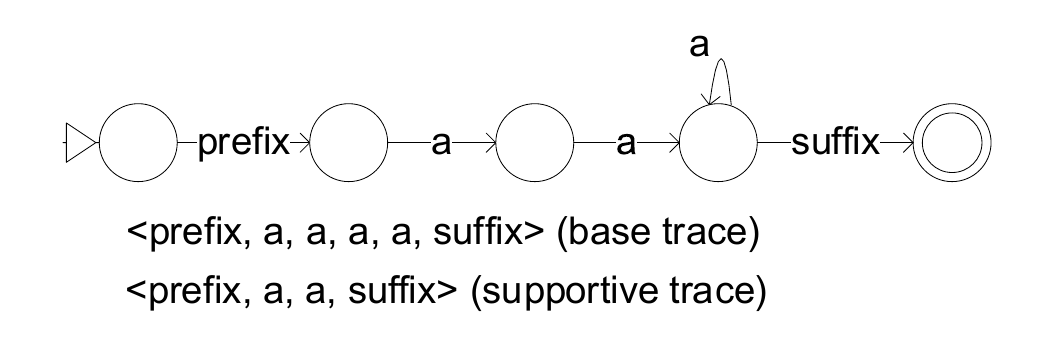}
    \caption{Case 3 (for repeat count (RC) = 2)} 
    \label{fig:fig10c}
\end{subfigure} 
 \caption{Heuristic for a loop: three loop models for a given base trace and three different supportive traces}
 \label{fig:loopheuristic}
\end{figure}

\blue{\fig}~\ref{fig:loopheuristic} shows three different generalized FSA models for a loop and their corresponding representative traces (a set of traces are used as evidence to support generalization). Given a base trace \textit{<prefix, a, a, a, a, suffix>} and \textit{RC} equal 2, there are three possible general models as shown in \blue{\fig}~\ref{fig:loopheuristic}. The ultimate general model depends on the other representative traces, called supportive traces. The FSA model shown in \blue{\fig}~\ref{fig:fig10a} will be the final general model if the supportive trace \textit{<prefix, suffix>} is found, i.e., it is present in the input traces. If the supportive trace \textit{<prefix, suffix>} can not be found, we search for a new supportive trace of the pattern \textit{<prefix, a, suffix>} by linearly increasing $ST$ - the number of loop operation (in this example $a$) in the supportive trace (in this case the general model will be as shown in \blue{\fig}~\ref{fig:fig10b}), while $BT-ST>RC$ and $BT$ is the number of loop operation in the base trace. In this example, the value of $BT-ST$ reaches 2 for case 3 (see \blue{\fig}~\ref{fig:fig10c}) and if the trace \textit{<prefix, a, a, a, suffix>} cannot be found in the input traces, we treat \textit{<prefix, a, a, a, a, suffix>} as just a sequence of transition without any generalization (iteration). We do not further proceed to consider \textit{<prefix, a, a, a, suffix>} because such a supportive trace will make the value of $BT-ST$ less than $RC$, i.e., 1.
\begin{figure}[htbp]
\centering
\begin{subfigure}{1\columnwidth}
 \centering
    \includegraphics[width=.55\textwidth]{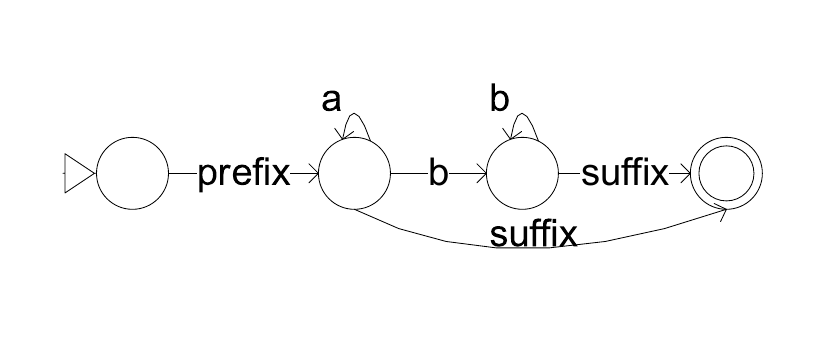}
    \caption{A reference model}
    \label{fig:fig11a}
\end{subfigure}
\par\bigskip 
\begin{subfigure}{1\columnwidth}
 \centering
    \includegraphics[width=1\textwidth]{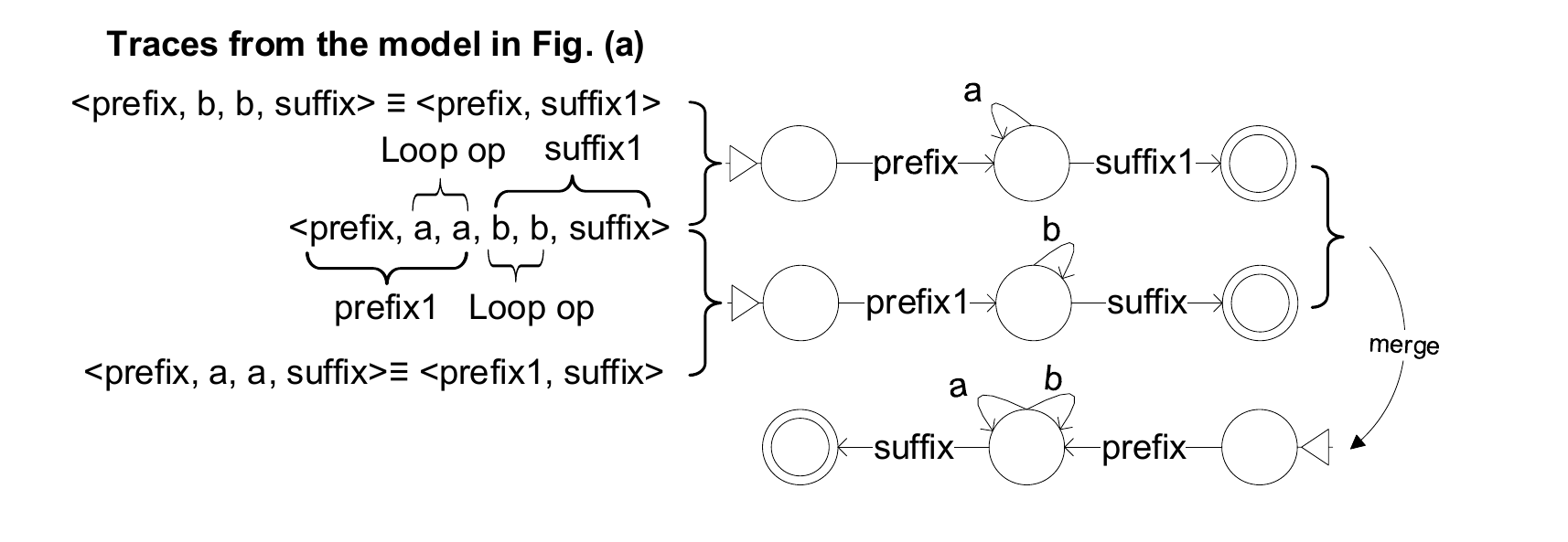}
    \caption{Applying loop heuristic causing over-generalization (see the reference model (a) and the inferred model}
    \label{fig:fig11b}
\end{subfigure}

\begin{subfigure}{1\columnwidth}
 \centering
    \includegraphics[width=1\textwidth]{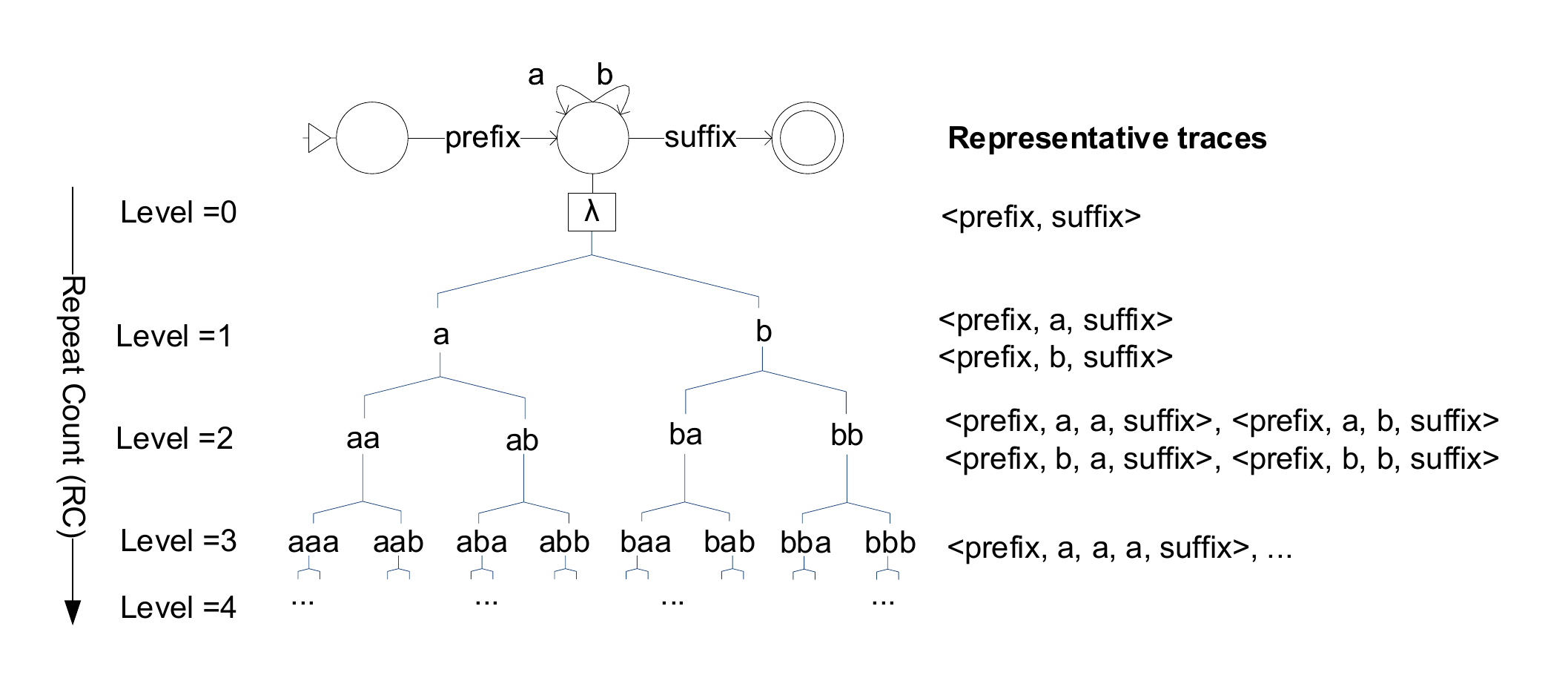}
    \caption{Example of minimum representative traces for a multiple-loop (2 loops in a same state); required representative traces increase with the value of repeat count/level number} 
    \label{fig:fig11c}
\end{subfigure} 
 \caption{Heuristic for a multiple-loop}
 \label{fig:multiloop}
\end{figure}

\subsubsection{Multi-loop}
A multi-loop is a special case of a (single) loop where a state has two or more loops. In this case, we need additional representative traces as evidence to support valid generalization. Let consider three traces <prefix, a, a, b, b, suffix>, <prefix, b, b, suffix> and <prefix, a, a, suffix> corresponding to the reference model shown in Fig.~\ref{fig:fig11a}. If we apply the loop algorithm in these traces, it will infer the multi-loop FSA shown in Fig.~\ref{fig:fig11b} which is an over-generalization with respect to Fig.~\ref{fig:fig11a}. To overcome this over-generalization we need additional representative traces as evidence to decide multi-loop.

If $l$ is a number of loops in a state, then the minimum number of required representative traces for a given $RC$ value is $\sum_{i=0}^{RC} l^i$.
Fig.~\ref{fig:fig11c} shows the representative traces for two loops example. The representative traces for $RC$ value 2 are all the corresponding traces from level 0 to 2. As the $RC$ value increases, the number of representative traces increases and so the evidence which makes generalization more precise. For example, when the above three traces lead generalization to a multi-loop, we further verify it by checking additional representative traces for $RC$ 2 (i.e., level = 0 to 2),  it will be seven representative traces as presented in Fig.~\ref{fig:fig11c}. Among these seven traces, the trace <prefix, b, a, suffix> cannot exist in the input traces for the model in Fig.~\ref{fig:fig11a}. Thus, this approach will prevent the over-generalization.

\begin{figure}[htbp]
\centering
\begin{subfigure}{1\columnwidth}
 \centering
    \includegraphics[width=.42\textwidth]{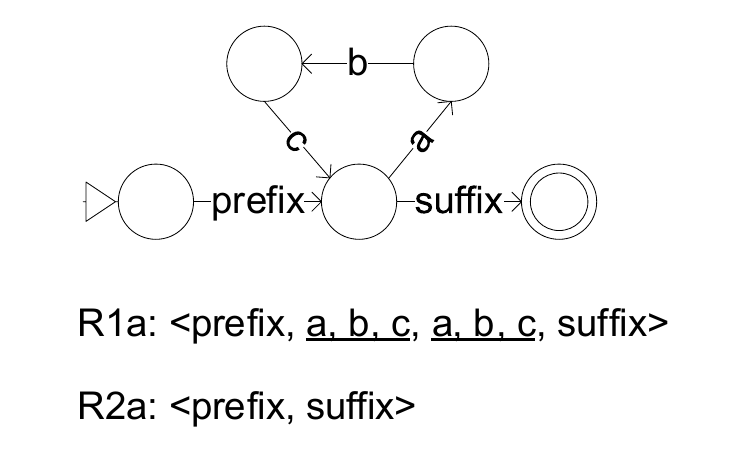}
    \caption{Case 1 model and its representative traces}
    \label{fig:fig12a}
\end{subfigure}
\hfill
\begin{subfigure}{1\columnwidth}
 \centering
    \includegraphics[width=.45\textwidth]{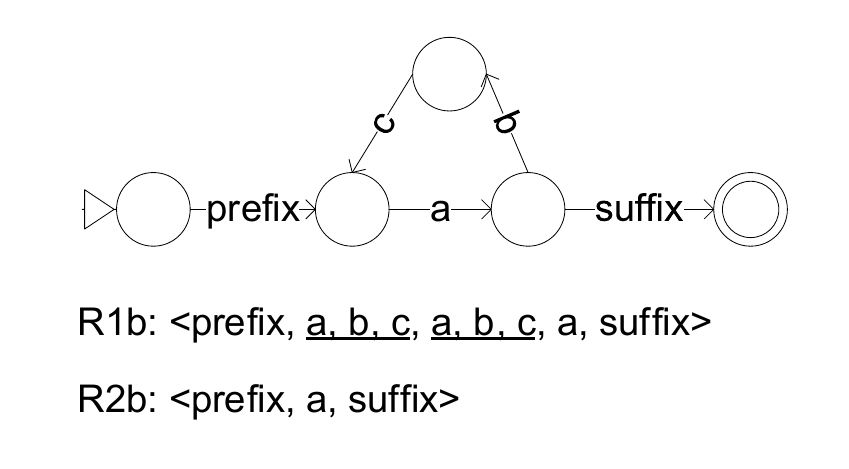}
    \caption{Case 2 model and its representative traces}
    \label{fig:fig12b}
\end{subfigure}

\begin{subfigure}{1\columnwidth}
 \centering
    \includegraphics[width=.55\textwidth]{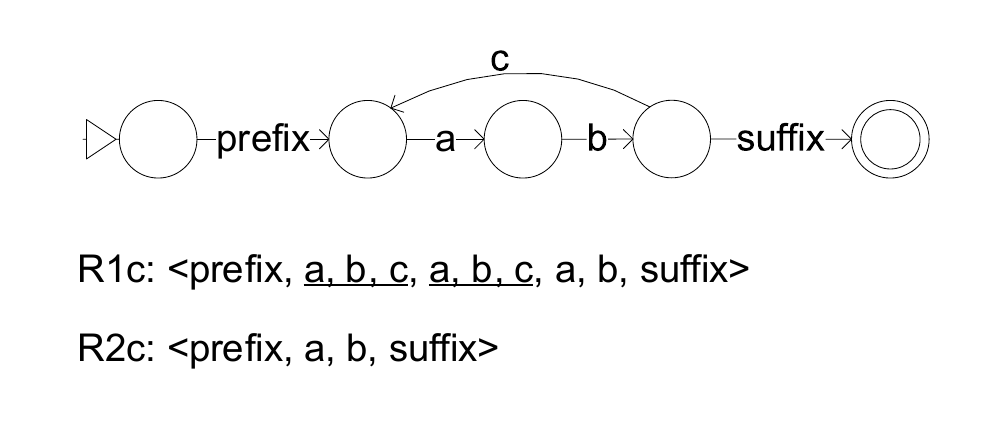}
    \caption{Case 3 model and its representative traces} 
    \label{fig:fig12c}
\end{subfigure} 
 \caption{Heuristic for a cycle: three different possible cases and their representative traces (for the repeat count (RC) value 2, i.e., twice repeat of events (underlined) are considered as sufficient evidence of a cycle)}
 \label{fig:cycleheuristic}
\end{figure}
\subsubsection{Cycle}
If a system model has a cycle, we expect that there should be a trace with repetition (at least twice, i.e., $RC$=2) of the cycle. For example, the trace \textit{<prefix, a, b, c, a, b, c, suffix>} shows a possibility of a cycle related to the sequence of operations \textit{<a, b, c>} which we can further confirm by checking whether the trace \textit{<prefix, suffix>} is present or not in the trace. If present, the generalization would be as shown in \blue{\fig}~\ref{fig:fig12a}. The presence of subsequent cycle operations before the suffix, in this example {a} and {a, b}, require a slight variation of the generalization as shown in \blue{\fig}~\ref{fig:fig12b} and \blue{\fig}~\ref{fig:fig12c}, respectively. 

\begin{figure}[htb]
\centering
\begin{subfigure}{1\columnwidth}
 \centering
    \includegraphics[width=.45\textwidth]{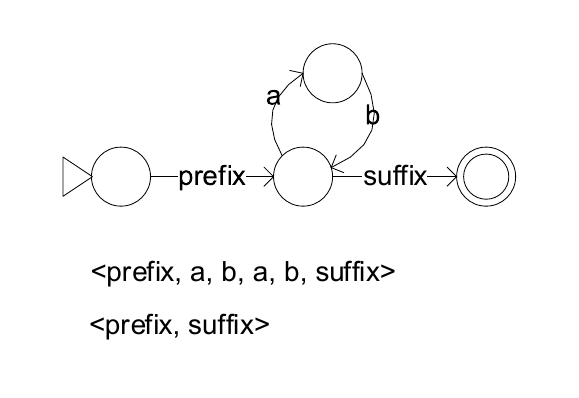}
    \caption{Case 1 model and its representative traces}
    \label{fig:fig13a}
\end{subfigure}
\begin{subfigure}{1\columnwidth}
 \centering
    \includegraphics[width=.48\textwidth]{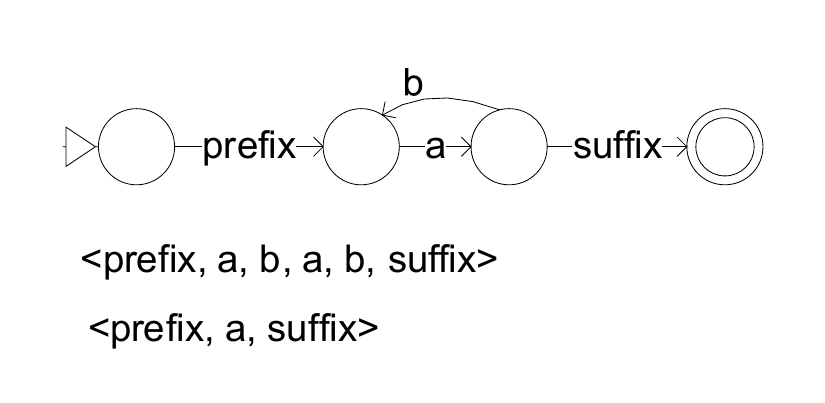}
    \caption{Case 2 model and its representative traces} 
    \label{fig:fig13b}
\end{subfigure} 
 \caption{Heuristic for a 2-cycle: two different model cases and their representative traces}
 \label{fig:cycleSpecialCase}
\end{figure}

An extra checking is required if the number of operations in the cycle is 2, called a \emph{2-cycle}. For example, the trace \textit{<prefix, a, b, suffix>} is valid for both models in \blue{\fig}~\ref{fig:cycleSpecialCase}. However, these models are not equivalent. To precisely identify the final model, we need additional representative traces to check further. In this case, for instance, if the trace \textit{<prefix, suffix>} is found as a valid trace (i.e., it exits in the input traces), we take the model shown in \blue{\fig}~\ref{fig:fig13a} as our generalized model. If the trace \textit{<prefix, a, b, suffix>} is found as a valid trace, our generalized model would be as shown in \blue{\fig}~\ref{fig:fig13b}.

\begin{figure}[htb]
\centering
\begin{subfigure}{1\columnwidth}
 \centering
    \includegraphics[width=.6\textwidth]{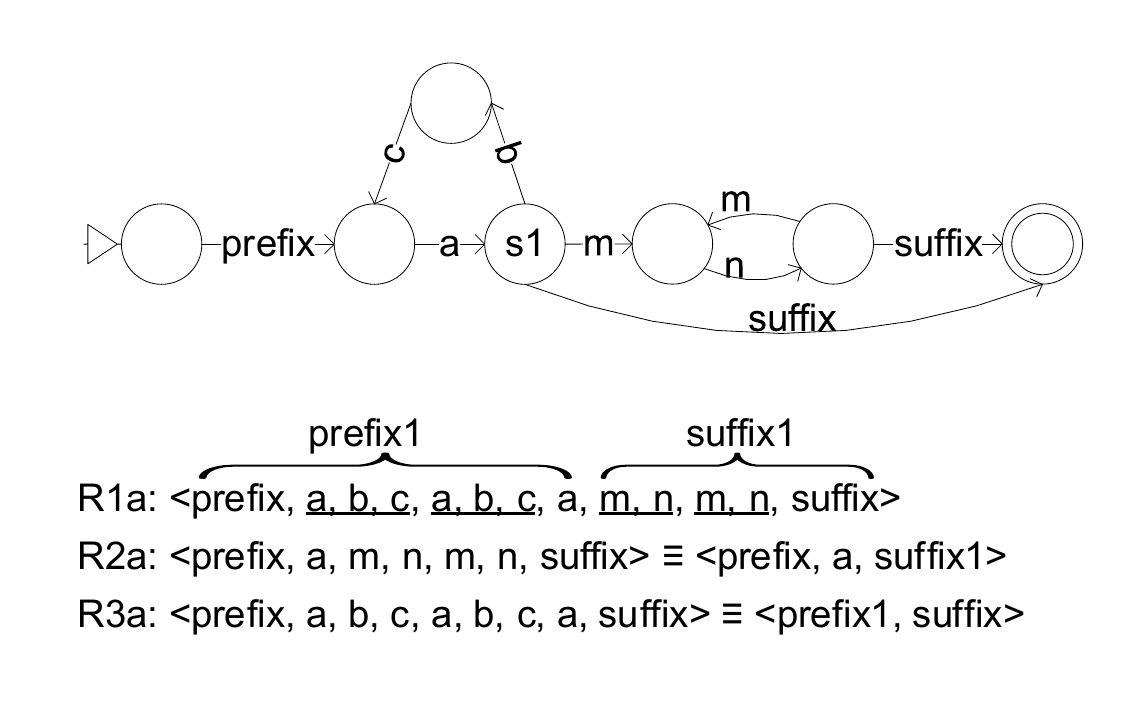}
    \caption{A model which is not a `Nested Cycle' but all its traces are accepted by the Nested Cycle model in Fig. (b)}
    \label{fig:fig14a}
\end{subfigure}
\begin{subfigure}{1\columnwidth}
 \centering
    \includegraphics[width=.6\textwidth]{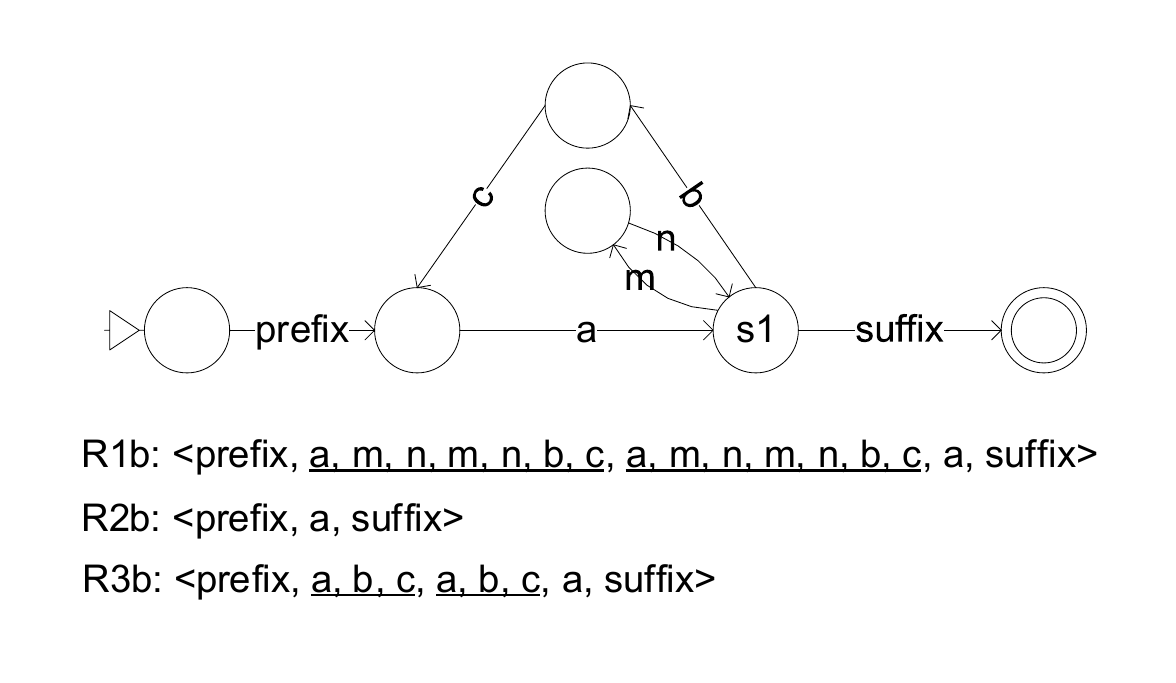}
    \caption{A `Nested Cycle' model and its representative traces} 
    \label{fig:fig14b}
\end{subfigure} 
 \caption{Heuristic of a Nested Cycle}
 \label{fig:nestedcycleheuristic}
\end{figure}

\subsubsection{Nested cycle}
A nested cycle is defined as  a cycle that contains other cycle(s). Let us consider the traces R1a, R1b and R1c corresponding to the reference model in \blue{\fig}~\ref{fig:fig14a}. If we apply the cycle heuristic as explained above in the trace \textit{<prefix, a, b, c, a, b, c, a, m, n, m, n, suffix>}, it results in a generalized model as shown in \blue{\fig}~\ref{fig:fig14b}, which is an \blue{over-generalization}~of the model in \blue{\fig}~\ref{fig:fig14a}. This \blue{over-generalization}~occurs because of making a cycle (for m, n) in a state (here, s1) which is already a part of another cycle (a, b, c). More specifically, a trace like R1b \textit{<prefix, a, m, n, m, n, b, c, a, m, n, m, n, b, c, a, suffix>} is acceptable by the model in \blue{\fig}~\ref{fig:fig14b}, but not acceptable by the model in \blue{\fig}~\ref{fig:fig14a}. 
For a model with a nested cycle as shown in \blue{\fig}~\ref{fig:fig14b}, we expect a trace R1b \textit{<prefix, a, m, n, m, n, b, c, a, m, n, m, n, b, c, a, suffix>} should exist in the input traces. First we find the largest repeated substring (as underlined in the trace R1b) that indicates cycle. Trace R2b provides further support for the existence of cycle. In the same way, iteratively, we check the existence of a repeated substring in a substring.

\subsubsection{Loop within cycle}
\begin{figure}[htb]
\centering
\begin{subfigure}{1\columnwidth}
 \centering
    \includegraphics[width=.6\textwidth]{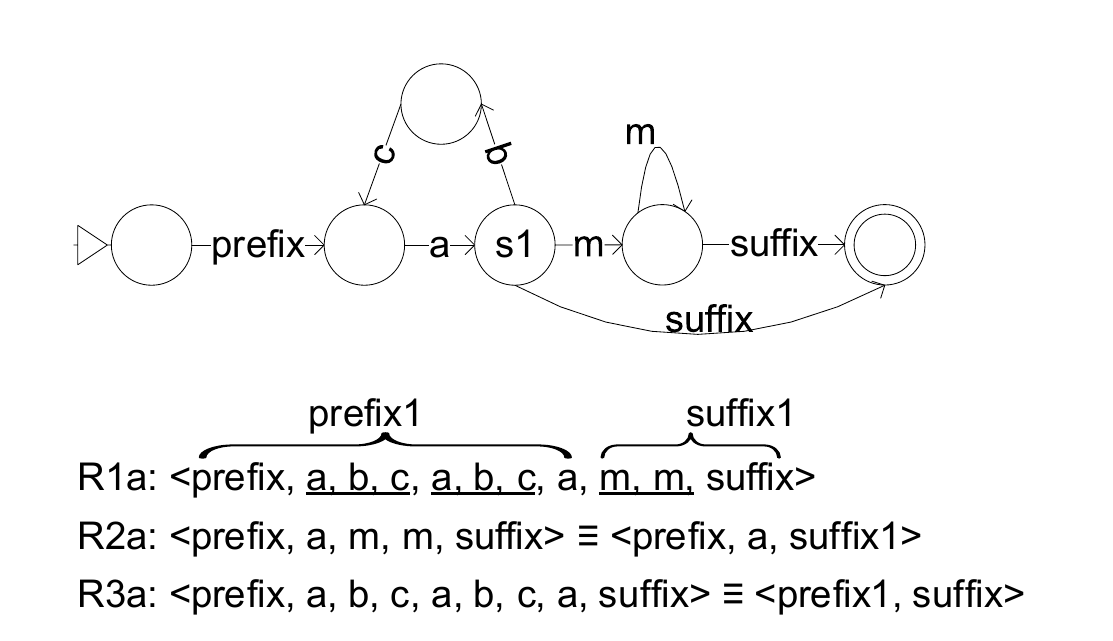}
    \caption{A model which is not a `Loop within Cycle' but all its traces are accepted by the Loop within Cycle model in Fig. (b)}
    \label{fig:fig15a}
\end{subfigure}
\begin{subfigure}{1\columnwidth}
 \centering
    \includegraphics[width=.55\textwidth]{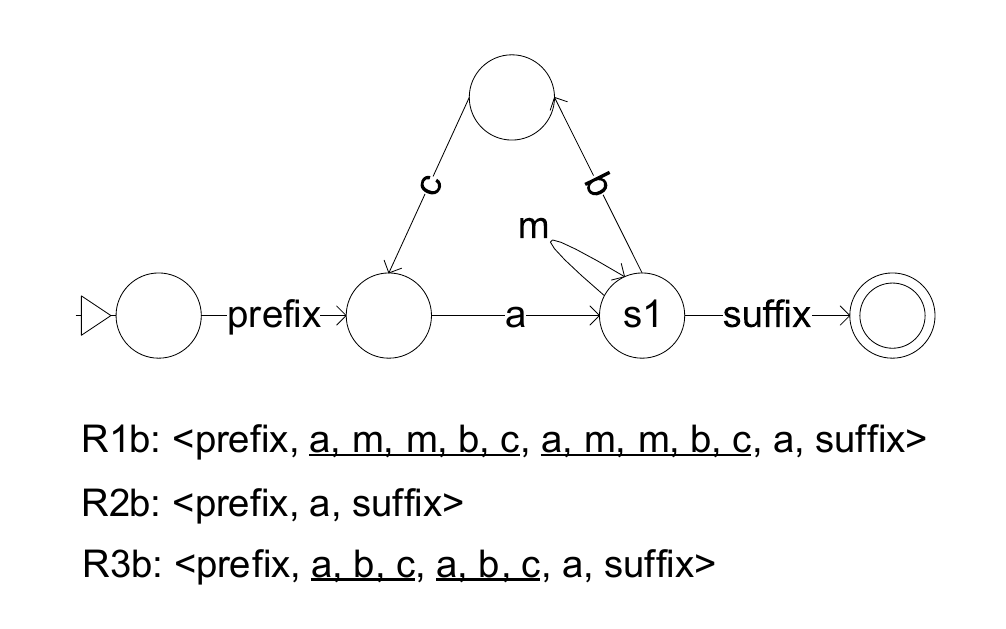}
    \caption{A `Loop within Cycle' model and its representative traces} 
    \label{fig:fig15b}
\end{subfigure} 
 \caption{Heuristic for a Loop within Cycle}
 \label{fig:loopwithincycle}
\end{figure}

\blue{\fig}~\ref{fig:loopwithincycle} present a case of loops(s) within a cycle. We address this case in a similar way to the nested cycle. \blue{\fig}~\ref{fig:fig15a} presents how we overcome over-generalization that is caused by applying loop and cycle heuristics as explained above without further consideration. \blue{\fig}~\ref{fig:fig15b} presents an example of a loop within cycle model and the required representative traces to infer that model. Both traces R1a and R1b are valid for the model in \purple{Fig.~\ref{fig:fig15b} } but for the model in Fig.~\ref{fig:fig15a} only the trace R1a is valid. For a loop within cycle model as shown in Fig.~\ref{fig:fig15b}, we expect a trace like R1b \textit{<prefix, a, m, m, b, c, a, m, m, b, c, a, suffix>} should exist in the input traces. First we find the largest repeated substring (in this example <a,m,m,b,c> as underlined in the trace R1b) that indicates cycle. Trace R2b provides further support for the existence of cycle. In the same way, iteratively, we check the existence of a repeated substring (here $m$) in a substring (in this example <a,m,m,b,c>).

\subsection{Minimization}
The generalization process results a set of deterministic FSAs for given interaction traces. To create a concise model, we connect those FSAs with an initial state using an empty ($\varepsilon$) transition which makes the model as a non-deterministic FSA, i.e., there exist more than one transition from the initial state for an empty transition.
We employ the Rabin–Scott powerset construction method~\cite{Rabin:1959} to convert the non-deterministic FSA to an equivalent deterministic FSA.
This conversion process may cause the generation of equivalent states, i.e., they have the same outgoing transitions, which means that they can be merged without changing the FSA properties. Indeed, merging equivalent states allows minimizing the FSA without changing the set of conversations it accepts. We use the Ullman-Hopcroft minimization algorithm \cite{hopcroft:2001} for this purpose.  

\subsection{Algorithm}

\begin{algorithm}[htbp]
	\caption{Inference Algorithm:  SpecMiner$(T, RC)$}\label{alg:InferenceAlgorithm}
	\begin{algorithmic}[1]
		\STATE \textit{Input:} set of traces, \emph{T} = $(t_1,..., t_n)$ where $t_i$ is a valid sequence of operations $<op_1, op_2,..., op_k>$ with $t_i[1]=op_1$ and $t_i[t_i.length]=op_k$, repeat count \emph{RC}$\geq$0
		\STATE \textit{Initialize:}  \emph{FSA} = \emph{start node}
		\STATE \textit{Output:} FSA

		\FOR {$t \in T$}
		\STATE $tFSA \leftarrow$ initialize $FSA$ with start node
		\FOR{$index \in \{1,\dots,|t|\} $}
		\IF {\textit{a sequence of operations $OP\in t$ consecutively repeated at least RC times 
				starting from $t[index]$}}
		\STATE Create Representative Traces (RT) based on $|OP|$, $t.prefix$, $t.cycOP$ and $t.suffix$ 
		
		\IF {$RT \subseteq T$}
		\STATE $cFSA \leftarrow$ SpecMiner($OP, RC$)   
		\STATE $tFSA \leftarrow$ add $cFSA$ as a cycle
		\STATE $index \leftarrow$  $index +|cycle|$
		\ELSE
		\STATE $tFSA \leftarrow$ add $cFSA$ as a sequence of transitions
		\STATE $index \leftarrow$  $index +|OP|$
		\ENDIF 
		\ELSE 
		\IF {\textit{the operation t[index] consecutively repeated at least RC times 
				starting from $t[index]$}}
		\IF {\textit{the current state of $tFSA$ contains a loop and the representative 
				traces for multiloop (created based on $RC$ value) $RT$ are found in $T$, i.e., 
				$RT \subseteq T$}}
		\STATE $tFSA \leftarrow $ add a loop transition with name $t[index]$
		\ENDIF
\algstore{myalg}
\end{algorithmic}
\end{algorithm}

\begin{algorithm}                     
\begin{algorithmic} [1]                   
\algrestore{myalg}
		\IF {\textit{the current state of $tFSA$ contains a cycle and the 
				representative traces for loop within cycle are valid}}
		\STATE $tFSA \leftarrow $ add a transition followed by a loop with name $t[index]$
		\ENDIF

		\IF {\textit{the representative traces for loop are valid}}
		\STATE $tFSA \leftarrow $ add the operation $t[index]$ as a sequence of transitions and loop
		\ELSE
		\STATE $tFSA \leftarrow $ add the operation $t[index]$ as a sequence of transitions 
		\ENDIF
		\STATE $index \leftarrow $ $index+ |loop|$
		\ELSE
		\STATE $tFSA \leftarrow$ create a new state with a transition name $t[index]$
		\STATE $index \leftarrow$  $index +1$
		\ENDIF
		\ENDIF
		\ENDFOR
		\ENDFOR
		\STATE $FSA \leftarrow $ ConvertNFAtoDFA($tFSA$)
		\STATE $FSA \leftarrow $ Minimize($FSA$)
		\STATE Return $FSA$	
		
	\end{algorithmic}
\end{algorithm}

Based on the heuristics and steps discussed in the previous section, this section presents the algorithm to infer a FSA based system model  from a set of system interaction traces with the given repeat count value (see Algorithm~\ref{alg:InferenceAlgorithm}). For each of the input traces (line 4), starting from the first operation of the trace to its end (line 6), the algorithm considers the operation sequence to identify whether the consecutive operations exhibit the behavior of a cycle (line 7) or a loop (line 18). If cycle, it applies the heuristics related to cycle as described in section \ref{subsec:heuric-driven generalization} (line 8 to 16). The algorithm first identifies repeated operations set, prefix and suffix, and accordingly creates representative traces (line 7 and 8). If these representative traces found valid (line 9), the algorithm is called recursively (line 10) by sending the repeated operation set as a trace to deal with the cases of nested cycle and/or loop within cycle. The resulted cFSA could be a simple FSA with a sequence of transitions (in case no inner loop or cycle) or a complex FSA (i.e., contains cycle and/or loop). Based on the prefix and suffix values, we incorporate this cFSA into tFSA as a cycle (line 11) (see \blue{\fig}~\ref{fig:cycleheuristic} for example). Then, the index of the trace is incremented to the position of operation that will be considered next (line 12). If the representative traces found invalid, we incorporate cFSA into tFSA as a sequence of transitions without cycle (line 14).

If an indication of loop is shown (line 18), we create and validate  the representative traces to identify whether it is a case of multiloop (line 19), loop within cycle (line 22), simple loop (line 25), or not loop (line 27), and generalize the trace accordingly. If the current position does not indicate a cycle or loop, a new state and a transition corresponding to the current operation are added to the tFSA. Once the generalization is complete for all traces, we  convert the final FSA (line 38) to \purple{Deterministic Finite Automata } (DFA) and minimize it (line 39).



\section{Evaluation}
\label{sec:evaluation}
We have empirically evaluated the effectiveness of our approach by comparing the quality of the models learned using SpecMiner with the models learned using kTail and Synoptic. In the following subsections, we describe the subject systems, experiment design and subject system datasets, and then present the results of our experiments.

\subsection{Subject systems}
We have applied SpecMiner to four different systems. In all cases, the input trace has been collected from ground-truth model using path-coverage and state-coverage strategies. All possible paths of ground-truth model have been followed at least once in path-coverage strategy and all states have been covered at least once in state-coverage strategy.
\begin{itemize}
	\item 
	\textbf{Concurrent Versions System (CVS)} is a free software client-server revision control system in the field of software development to keep track of all work and all changes in a set of files, and allows several developers (potentially widely separated in space and time) to collaborate. CVS data-set contains 20 different interaction traces with average 12 messages in each trace. 

	\item
	\textbf{Amazon-ec2} is a commercial web service from Amazon's Web Services (AWS) that lets customers ``rent'' computing resources from the Elastic Compute Cloud (EC2). It provides storage, processing, and web services to customers. Amazon-ec2 data-set contains 76 different interaction traces with average 9 messages in each trace. 
	\item 
	\textbf{StringTokenizer} is a Java library (\emph{java.util.StringTokenizer}) for breaking a string into tokens. It is a simple way to break string and doesn't provide the facility to differentiate numbers, quoted strings, identifiers etc. StringTokenizer data-set contains 39 different interaction traces with average 4 messages in each trace. 

	\item
	\textbf{ZipOutputStream} is a Java library (\emph{java.util.zip.ZipOutputStream}) for writing files in the ZIP file format, which includes support for both compressed and uncompressed entries. ZipOutputStream data-set contains 799 different interaction traces with average 11 messages in each trace. 
	
\end{itemize}

\blue{Table \ref{tab:datasetDistribution} shows the distribution of message types in each dataset.}

   \begin{table}[hbt]
	\centering
	\caption{Distributions of Messages in each dataset}
	\label{tab:datasetDistribution}
\scalebox{0.8}{
		\begin{tabular}{p{1.6cm}p{4cm}p{2cm}p{2.4cm}p{2cm}}
	\hline
	
	 & CVS client & Amazon-ec2 & String-Tokenizer & ZipOutput-Stream
	    \\\hline
	    \hline
	  No. of Traces & 20 & 76 & 39 & 799 \\\hline
	  No. of Messages & 258 & 706 & 181 & 8975 \\\hline
	  Message Type (Distribution) & setfiletype (9.69\%), connect (7.75\%), retrievefile (5.43\%), rename (5.97\%), initialise (7.75\%), storefile (9.30\%), makedir (1.16\%), listfiles (8.14\%), changedir (9.69\%), listnames (4.26\%), appendfile (3.87\%), appendfile (1.16\%), logout (7.75\%), delete (1.55\%), login (7.75\%), disconnect (7.75\%) & startInstance (25.78\%), rebootInstance (16.99\%), stopInstance (27.48\%) & hasMoreTokens (39.23\%), nextToken (39.23\%), init (21.54\%) &  ZipOutput- Stream (8.90\%), closeEntry (20.83\%), close (8.90\%), putNext- Entry (43.32\%), write (18.05\%)

	   \\\hline
	   
	   \hline
		\end{tabular}}
\end{table}

\subsection{Evaluation setup}
In this section, we describe the ground-truth models of our four subject systems, evaluation metrics and experimental design.

\subsubsection{Ground-truth models}
Evaluating the quality of the inferred models requires a set of ground-truth models that represent the systems' legal behaviors. To this end, we have used the ground-truth models of three subject systems such as CVS Client, StringTokenizer and ZipOutputStream that were manually extracted as part of related work~\cite{lo:2006b,dsm:david:lo}. The ground truth model of Amazon-EC2 is constructed manually by analyzing its service interface.

\subsubsection{Evaluation metrics}
\blue{For a given subject system, we use two standard evaluation metrics, Precision and Recall, to evaluate the effectiveness of SpecMiner. We also measure the time required to infer the behavioral model from the interaction traces. \textit{True positive} is the number of behavioral patterns that are accepted by both the ground-truth model and the mined model. \textit{False positive} is the number of behavioral patterns that are accepted by the mined model but rejected by the ground-truth model. \textit{False negative} is the number of behavioral patterns that are rejected by the mined model but accepted by the ground-truth model.}    

\subsubsection{Experimental design}
For each subject system and its dataset, we have inferred four models, each with traditional kTail with k=1 and k=2, Synoptic~\cite{schneider:2010, beschastnikh:2011} (which is an improved kSteering approach), and our approach (with RC value 2 as this value is high enough to protect spurious merges for the considered four systems). We do not report results of kTail for $k>2$ because those $k$ values led to fewer merges, lowering the recall without notable precision improvement~\purple{\cite{lo:2009, biermann:1972}}.

In the experiments, for each subject system, we perform three steps: (1) generate input traces from the ground truth model with path coverage~\cite{pathcov} where each transition is visited maximum twice, (2) run the four inference algorithms on those input traces, and (3) measure the precision and recall of the inferred models. 


\subsection{Results}
We compare the effectiveness and efficiency of our approach with kTail (k=1 and k=2) and Synoptic by computing the precision and recall for the resulting automata and measuring the time required to generate the models. Table \ref{tab:empiricalresult} shows the results of our experiments.

\begin{table*}[tbh]
	\centering
	\caption{Empirical Result}
	\label{tab:empiricalresult}
\scalebox{0.88}{
	\begin{tabular}{p{0.18\textwidth}|p{0.06\textwidth}p{0.06\textwidth}p{0.04\textwidth}|p{0.02\textwidth}p{0.06\textwidth}p{0.04\textwidth}|p{0.02\textwidth}p{0.02\textwidth}p{0.04\textwidth}|p{0.03\textwidth}p{0.03\textwidth}p{0.06\textwidth}}
		\hline
		\multirow{2}{*}{\parbox{2cm}{\textbf{Approaches}}}
		& \multicolumn{3}{|p{0.12\textwidth}}{\textbf{CVS client}} 
		&   \multicolumn{3}{|p{0.12\textwidth}}{\textbf{Amazon-ec2}} 
		&   \multicolumn{3}{|p{0.09\textwidth}}{\textbf{String Tokenizer}} 
		&    \multicolumn{3}{|p{.12\textwidth}}{\textbf{Zip- Output- Stream}}\\
		\cline{2-13} & P & R & T & P & R  & T & P & R & T & P & R & T \\ \hline
		\hline
		kTail (k=1) &0.002 & 0.528 & 68 & 1 & 1 & 147 & 1 & 1 & 19 & 1 & 1 & 47  \\
		kTail (k=2) & 1 & 0.339 & 90 & 1 & 0.346 & 120 & 1 & 1 & 15 & 1 & 1 & 34  \\ 
		Synoptic & 0.244 & 1 & 396 & 1 & 1 & 373 & 1 & 1 & 611 & 1 & 1 & 337  \\
		Our approach &  1 & 1 & 47 & 1 & 1 & 65 & 1 & 1 & 43 & 1 & 1 & 99  \\ \hline
		
		\hline
		\multicolumn{13}{l}{\textsuperscript{*}Note: P is precision, R is recall, and T is time in milliseconds.}
	\end{tabular}}
\end{table*}
\purple{The results show that our approach can infer models of all four systems with 100\% precision and recall. In particular, our approach outperforms kTail and Synoptic for CVS client system as 
kTail (k=2) under-generalizes and Synoptic over-generalizes the infer model. } 
We observe that the highest difference in precision when experimenting with the CVS client system. In this the case, given the best recall (i.e., 1), the precision of the inferred model raises from 0.244 in the case of model inferred by Synoptic, to 1 in the case of model inferred by our approach. It is worth noticing that in this study our approach always generates models with perfect precision (i.e., 1).

\purple{While our approach, Synoptic and kTail with k=1 infer model with 100\% precision and recall for Amazon-ec2, but kTail with k=2 under-generalize it's infer model. However, our approach takes less time to infer model compare to Synoptic and kTail.

Unlike CVS client, for StringTokenizer and ZipOutputStream 
all four approaches infer models with 100\% precision and recall. However our approach takes less time compare to Synoptic but little more time compare to kTail.}




\section{Related Work}
\label{sec:related work}
There have been numerous works in the area of specification or behavior mining. They can be classified into two groups, depending on what the mined specifications represent: order of events (e.g.,~\cite{lo:2009}) and invariants on values of program variables (e.g.,~\cite{beschastnikh:2011}).

\subsection{Mining models of ordered events}
\purple{In this section, we review the existing literature that deduces specification or behavior models as ordered of events. Those models can be classified as Finite State Automata (FSA) and non-FSA based models based on their representations}.
\subsubsection{FSA-based Models}
There are many simple FSA learners~\cite{biermann:1972, Mariani2007, Dallmeier:2006, Ammons:2002, raffelt2006learnlib, Walkinshaw:2007, Antunes:2011, Bertolino:2009, schneider:2010}. The approach ranges from passive (i.e., only analyse execution traces) to active (i.e., ask queries to users or generate test cases such as~\cite{raffelt2006learnlib, Walkinshaw:2007, Bertolino:2009}). These learners also analyse different kind of execution traces, from standard program execution traces to network traces~\cite{Antunes:2011}, web service invocations~\cite{Bertolino:2009}, and system logs~\cite{schneider:2010}.
The kTail algorithm \cite{biermann:1972} serves as a basis for many approaches for mining FSA models from invocation traces~\cite{cook:1998,lo:2006a,lo:2006b,lo:2009,lorenzoli:2008,reiss:2001}. These approaches extend kTail to (1) improve its precision or recall~\cite{cook:1998,lo:2009,reiss:2001}, (2) incrementally generate a FSA~\cite{r6:Mariani:2011}, (3) build larger frameworks with kTail as the inference algorithm~\cite{lamprier2015care,lo:2006b}, and (4) enhance the models with information about invocation probabilities~\cite{lo:2006b} and program state and method parameters~\cite{lorenzoli:2008}. These approaches extend kTail mainly by either considering (1) trace invariants \cite{lo:2009} or (2) values of method parameters \cite{lorenzoli:2008, r4:4700316, krka:2014}, in addition to kTail. As all of these approaches have adopted the kTail algorithm with some extensions,  the key limitation of kTail's principle, i.e., inter-trace state merging, also applies to these approaches.

Synoptic \cite{schneider:2010, beschastnikh:2011} uses the CEGAR \cite{clarke:2000} approach to create a coarse initial model (which is extremely overgeneralized), and then refine it using counter examples that falsify temporal invariants. Synoptic represents the model as a graph, it can easily be converted to an equivalent FSA model. Meanwhile, InvariMint \cite{beschastnikh:2013} presents a declarative specification language for expressing model-inference algorithms, and improves the efficiency of the algorithms, but not their precision or recall. 
 
\subsubsection{Non FSA-based Models}

\blue{Although FSA is a popular representation of behavioral models, many other techniques infer behavioral models in other forms, such as sequence diagrams, frequent patterns, and temporal properties. Even though the different techniques infer behavioral models in different forms, yet the ultimate goal of these techniques is to capture the system behaviors from the traces.}


\blue{A group of existing techniques infers sequence diagrams and patterns from the traces as behavioral models~\cite{Hosking:1996, McGavin:2006:VET, Briand:2006:TRE, Lo:2007:MMS, Lo:2008:MST, el2002interaction, Lo:2007:EMI, Safyallah:2006:DAS}. Some of them infer behavioral models from the traces in the form of sequence diagrams~\cite{Hosking:1996, McGavin:2006:VET}. These techniques just describe the input traces as sequence diagrams. A few of them are capable of inferring loops~\cite{Briand:2006:TRE}, while others infer frequent patterns that appear repeatedly in the traces~\cite{Lo:2007:MMS, Lo:2008:MST, el2002interaction, Lo:2007:EMI, Safyallah:2006:DAS}. Another group of existing techniques infers temporal properties as behavioral models from the traces. The technique presented by Yang et al. mines important future temporal rules having satisfaction rates defined by the user~\cite{Yang:2006:PMT}. The techniques presented by Lo et al. mine future-rules of arbitrary lengths~\cite{Lo:2008:MTR:1400155} and past-rules of arbitrary lengths~\cite{Lo:2008:MPT}.}


\blue{In this work, an FSA-based behavioral model is inferred from the traces. In general, FSA describes the relationships between events more effectively comparing to the temporal properties and frequent patterns.  Temporal rules describe only partial relations between events and cannot completely capture loops and branches, whereas frequent patterns describe only linear relationships among events.}  

\subsection{Mining invariants on values of program variables}
Different from the goal of this work, there are several techniques that mine behavioural models in the form of invariants on values of program variables. These models usually represent aspects that complement the information described by FSA. The most popular example is Daikon~\cite{ernst:2001}, a mining approach to learn boolean expressions that describe likely invariants on values of program variables. Similar to Daikon, the Contractor technique~\cite{de:2012} creates models based solely on the inferred state invariants. Contractor++~\cite{krka:2014} addresses the limitations of Contractor when applied to dynamically inferred, as opposed to manually specified, state invariants. TEMI \cite{krka:2014} infers models from invariants and then enhances them with execution traces. The first phase is conceptually similar to Contractor++, capturing all invocation sequences of a system allowed by the invariants. The second phase promotes transitions observed in the traces.



\section{Threats to validity}
\label{sect:threat}
In this section, we outline the possible threats to our evaluation's validity and discuss our mitigation strategies.

The ground-truth models were manually constructed. This may make them biased to the specifier's expertise. To mitigate this threat, we used the publicly available ground-truths from other researchers.

The selection of libraries may bias the evaluation results. For instance, models of a well-known library may not be representative of dynamically inferred models in general. Thus, we selected libraries from different domains and of various sizes. 

There are multiple ways of comparing the generated models with the corresponding ground-truth models (e.g., recall and precision of the simulated traces vs. graph comparison), which could potentially yield different results. Our mitigation strategy was to use metrics that have been proposed and adopted by the research community~\cite{LO20122063,krka:2014}, and are consistent with our goal of comparing model behaviour, not structural similarity.


\section{Conclusions}
\label{sec:conclusion}
To improve accuracy of the inferred FSA-based behavior models of systems from their interaction traces, \purple{this paper proposes } a heuristic-driven technique to generalize the system interaction traces into behavior models. In particular, we start with intra-trace state merging based on a rich set of heuristics that consistently address different types of generalization cases in a FSA. \blue{The basis of our heuristics is the system behavior patterns present in the traces which provide us evidence for generalization. The consideration of these heuristics has allowed us to take into account the finer-grained behavioural relationships that exist in system interaction traces, leading to the generation of more accurate system behavioral models. Furthermore, we have systematically considered heuristics for all possible fundamental building blocks of FSA from simple cases such as loop, cycle, etc to complex cases such as nested loop, nested cycles, etc. Those cases could be used to deal with any further complex cases. } We have evaluated our technique on a number of systems from various domains. These empirical experiments have shown that our approach can significantly increase the precision of the mined models without loss of recall, thereby increasing the overall accuracy of the models mined. The experiments have also shown that our inference process only incurs a limited computation overhead. 

This work is significant as it provides a way to mine richer and more accurate system behavior models or specifications than existing approaches, increasing the utility of these models in system verification, validation and emulation. Furthermore, this work also lays the foundation for further work on  technique to mine  parametric system models and specifications.


\section*{Acknowledgements}
This work has been supported by the Australian Research
Council (ARC) and CA Technologies.



%
%
%

\bibliography{sigproc}

\end{document}